\documentclass[aps,prb,twocolumn,superscriptaddress]{revtex4-1}

\usepackage{mathtools}
\usepackage{lipsum}
\usepackage{amsmath,amssymb}
\usepackage{MnSymbol}
\usepackage{graphicx}
\usepackage{color}
\usepackage{hyperref}
\usepackage{url}
\usepackage{slashed}
\usepackage{slashed,bbm}
\usepackage{graphics,psfrag,epsfig}
\usepackage{dsfont}
\usepackage{enumerate}
\usepackage{pbox}
\usepackage{braket}
\usepackage[retainorgcmds]{IEEEtrantools}
\usepackage{xr}
\usepackage{gensymb}

\DeclareMathSizes{10}{10}{6}{4}

\graphicspath{{./figures/}}

\begin{document}
	
	\title{Rashba Sandwiches with Topological Superconducting Phases}
	
	\author{Yanick Volpez}
	\affiliation{Department of Physics, University of Basel, Klingelbergstrasse 82, CH-4056 Basel, Switzerland}
	
	\author{Daniel Loss}
	\affiliation{Department of Physics, University of Basel, Klingelbergstrasse 82, CH-4056 Basel, Switzerland}
	
	\author{Jelena Klinovaja}
	\affiliation{Department of Physics, University of Basel, Klingelbergstrasse 82, CH-4056 Basel, Switzerland}
\date{\today}	
	
	\begin{abstract}
We introduce a versatile heterostructure harboring various topological superconducting phases characterized by the presence of helical, chiral, or unidirectional edge states. Changing parameters, such as an effective Zeeman field or chemical potential, one can tune between these three topological phases in the same setup. Our model relies only on conventional non-topological ingredients. The bilayer setup consists of an $s$-wave superconductor sandwiched between two two-dimensional electron gas layers with strong Rashba spin-orbit interaction. The interplay between two different pairing mechanisms, proximity induced direct and crossed Andreev superconducting pairings, gives rise to multiple topological phases.  In particular, helical edge states occur if crossed Andreev superconducting pairing is dominant. In addition, an in-plane Zeeman field leads to a 2D gapless topological phase with unidirectional edge states, which were previously predicted to exist only in non-centrosymmetric superconductors. If the Zeeman field is tilted out of the plane, the system is in a topological phase hosting chiral edge states.
\end{abstract}
	\maketitle

\section{Introduction}
The discovery that certain properties of quantum states of matter can be captured in terms of topological invariants \cite{TKNN,KaneMeleZ2,Kitaev2009} immune to microscopic details of a system has triggered enormous interest in the exploration of topological phases of matter \cite{Hasan2010,Qi2011,Sato2017,Stern2016}. Currently, a great effort is put into the search for localized Majorana quasiparticles that are predicted to appear in one-dimensional topological superconductors (TSCs) \cite{Kitaev2000,Ivanov2001,Lutchyn2010,Oreg2010,Alicea2010,Potter2011,Chevallier2012,Klinovaja2012_3,Sticlet2012,Klinovaja2012_2,Halperin2012,NadjPerge2013,Klinovaja2013,Braunecker2013,Vazifeh2013}. Two-dimensional (2D) TSCs are the particle-hole symmetric analogues of the experimentally more extensively studied topological  insulators (TI)\cite{Pankratov1987,Bernevig2005,KaneMeleQSH,Koenig2007,Roth2009,Gusev2013,Gusev2014,Ma2015,Olshanetsky2015,Deacon2017,Liu2008,Knez2011,Suzuki2013,Charpentier2013,Knez2014,Suzuki2015,Mueller2015, Qu2015, Li2015,Du2015, Nichele2016,Couedo2016,Akiho2016,Nguyen2016,Mueller2017}. One-dimensional (1D) TSCs have been subject of intense experimental research \cite{Mourik2012,Das2012,Albrecht2016,Deng20122,Churchill2013,Finck2013,NajPerge2014,Ruby2015,Pawlak2016, Chang2015,Deng2016,Gazibegovic2017,Zhang2017,Kjaergaard2016,Shabani2016,Kjaergaard2017,Suominen2017,Nichele2017}, while 2D TSCs are not yet so well-developed experimentally. However, various setups hosting chiral \cite{Qi2010,Qi20102,Yoshida2016,Daido2016,Menard2017,Li2016,Rontynen2015,Nakosai2013,Alicea2011} and helical \cite{Fu2008,Sato2009,Qi2009,Fu2010,Trauzettel2011,Nakosai2012,Deng2012,Zhang2013,Parhizgar2016,Wang2016} superconducting edge states were proposed theoretically. In addition to gapped TSCs, there exist gapless TSCs, which are predicted to be realized in nodal superconductors with mixed singlet-triplet pairing and Rashba spin-orbit interaction (SOI) or in various heterostructures \cite{Wong2013,Hao2017,Daido2017,Schnyder2015,Schnyder2011,Sato2011,Sato2010,Tanaka2010,Meng2012,Deng2014,Huang2018}. The majority of these proposals involve topological insulators and/ or unconventional superconductors with $p$-, $d$-wave pairing symmetry.
\begin{figure}[b]
	\includegraphics[width=.7\columnwidth]{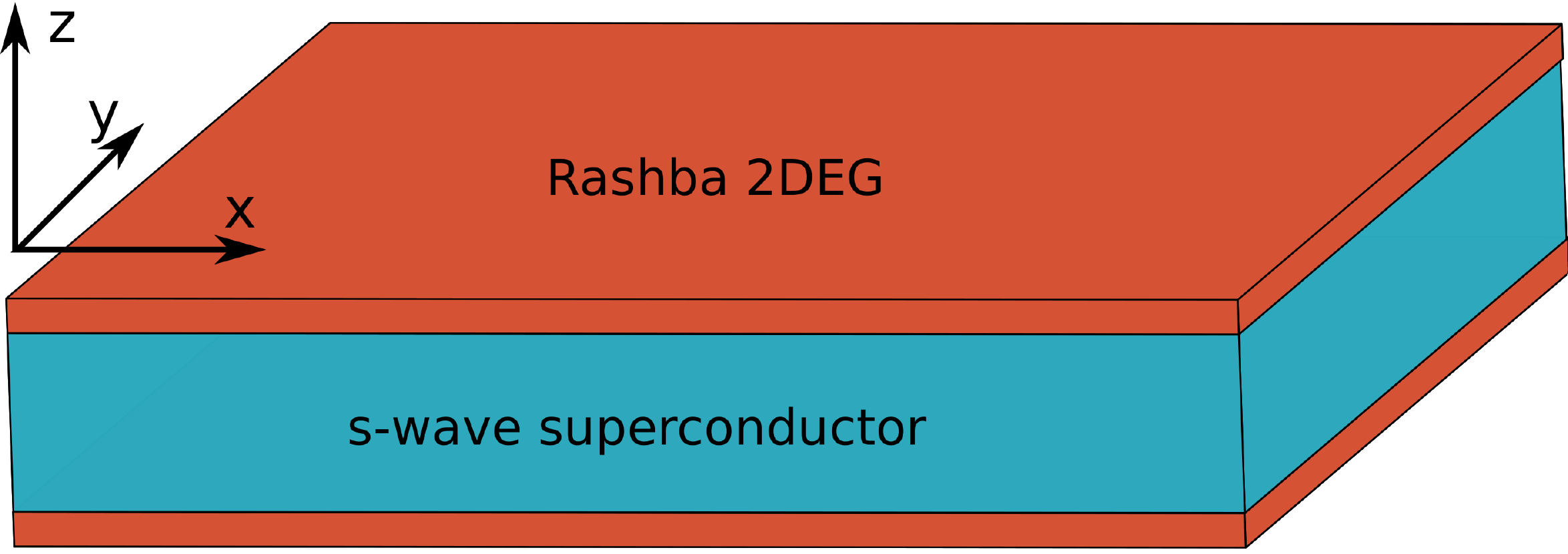}
	\caption{Sketch of the bilayer setup consisting of an $s$-wave superconductor (blue) sandwiched between two two-dimensional electron gas layers (red) with strong Rashba spin-orbit interaction.}
	\label{Setup}
\end{figure}

In this work, we propose a bilayer heterostructure which can be brought into all the 2D topological phases mentioned above without the need of including topological materials and/or unconventional superconductors (SCs). The setup is composed of only conventional components: an $s$-wave superconductor sandwiched between two two-dimensional electron gas (2DEG) layers with strong Rashba spin-orbit interaction (SOI) (see Fig.~\ref{Setup}). The proximity to the superconductor induces superconducting correlations in the 2DEG layers with direct and crossed Andreev pairings\cite{Byers1995,Choi2000,Deutscher2000,Lesovik2001,Recher2001,Yeyati2007,
Reeg2017_1,Reeg2017_2,ManishaRG,Schoenenberg2009,Heiblum2012,Tokyo}. In the former case, a Cooper pair tunnels into one layer, whereas, in the latter case, a Cooper pair splits and the electrons tunnel to opposite layers respectively. When crossed Andreev pairing is dominant, the system is in a gapped phase with a Kramers pair of helical edge states, {\it i.e.} it is a helical TSC. 

Interestingly, if an effective Zeeman field is introduced, {\it e.g.} due to externally applied magnetic fields or due to the ordering of magnetic impurities, the bilayer setup has the potential to realize either a chiral or a gapless TSC. If the Zeeman field lies in the plane, then, above a certain critical field strength, the system enters a gapless topological phase with unidirectional edge states. Unidirectional edge states, which are states that propagate in the same direction on opposite edges, appear on the edge orthogonal to the direction of the in-plane field \cite{Daido2017}. If the Zeeman field points out of the plane with an angle larger than a model parameter dependent threshold, the system enters again a fully gapped phase with chiral edge states. In contrast to the unidirectional states, the chiral edge states propagate in opposite directions on opposite edges and appear on all boundaries of the system.

To the best of our knowledge, none of the theoretical proposals for realizing 2D TSCs by solely including conventional non-topological ingredients was reported to be able to obtain all of the topological phases mentioned above.

The paper is organized as follows. In Section \ref{Model}, we introduce the effective model describing our setup and discuss its bulk properties in the absence of a Zeeman field. In Section \ref{HTSC}, we show in which parameter regimes the system is a helical TSC, characterize the spin and charge properties of the helical edge states, and derive the effective low-energy Hamiltonian for the edge states. In Section \ref{MagneticField}, we study the possible topological phases in the presence of a Zeeman field. We show that the helical TSC, protected by time-reversal symmetry, gets immediately destroyed, and the topological phases with  chiral and gapless edge states become accessible.

\section{Model} \label{Model}

 We consider a bilayer setup consisting of two 2DEG layers with strong Rashba SOI coupled to an $s$-wave superconductor (see Fig. \ref{Setup}). Each layer is characterized by the SOI strength $\alpha_{\tau}$, where we label the upper (lower) layer by the index $\tau=1$ $(\tau=\bar{1})$. In the following, we restrict the discussion to the case $\alpha_1>\alpha_{\bar{1}}>0$. The $z$ axis is normal to the layers and ${\bf k}=(k_x,k_y)$ is the in-plane momentum. The Hamiltonian describing the two uncoupled layers reads in momentum space as
\begin{align}\label{HamiltonianFreeLayer}
H_{\tau}= \sum_{\sigma, \sigma'} \int d^2 {\bf k} \ \psi^{\dagger}_{\tau \sigma,\mathbf{k}} h_{\tau \sigma \sigma'}({\bf k}) \psi_{\tau \sigma',\mathbf{k}}, 
\end{align}
where $h_{\tau \sigma \sigma'}({\bf k})= [\epsilon_k-\mu_{\tau} + \alpha_{\tau} {\bf g} \cdot \boldsymbol{\sigma}]_{\sigma \sigma'}$ with $\epsilon_k=\hbar^2 |{\bf k}|^2/2m$ and ${\bf g}=(k_y,-k_x,0)$. The field operator $\psi_{\tau \sigma,\mathbf{k}}$ annihilates an electron in the layer $\tau= \{1, \bar{1} \}$ with spin projection $\sigma= \{1, \bar{1} \}$ and momentum ${\bf k}$. In what follows, the chemical potentials $\mu_\tau$ are tuned to the spin-orbit energy of the respective layer $E_{so,\tau}=m\alpha_{\tau}^2/2\hbar^2$. 

The proximity-induced superconductivity opens gaps in the spectrum of the bilayer system and is responsible for the topological phase. Generally, there are two types of superconducting terms: direct and crossed Andreev pairing terms of strength $\Delta_D$ and $\Delta_C$, respectively. The direct (crossed Andreev) proximity-induced superconductivity induces coupling between two electrons from the same layer (from two different layers),
\begin{align}\label{HamiltonianSuperconductivity}
H_D &= \frac{\Delta_D}{2} \sum_{\substack{\tau \\ \sigma, \sigma'}} \int d^2 {\bf k} \ \Big(\psi^{\dagger}_{\tau \sigma,{\bf k}} [i \sigma_2]_{\sigma \sigma'} \psi^{\dagger}_{\tau \sigma',-{\bf k}} + \text{H.c.}\Big), \\
H_C &= \frac{\Delta_C}{2} \sum_{\substack{\tau \\ \sigma, \sigma'}} \int d^2 {\bf k} \ \Big( \psi^{\dagger}_{\tau \sigma,{\bf k}} [i \sigma_2]_{\sigma \sigma'} \psi^{\dagger}_{\bar{\tau}  \sigma',-{\bf k}} + \text{H.c.} \Big).
\end{align}
Without loss of generality we assume $\Delta_C$, $\Delta_D>0$ throughout this work. 

\begin{figure}
	\centering
	\includegraphics[width=.75\columnwidth]{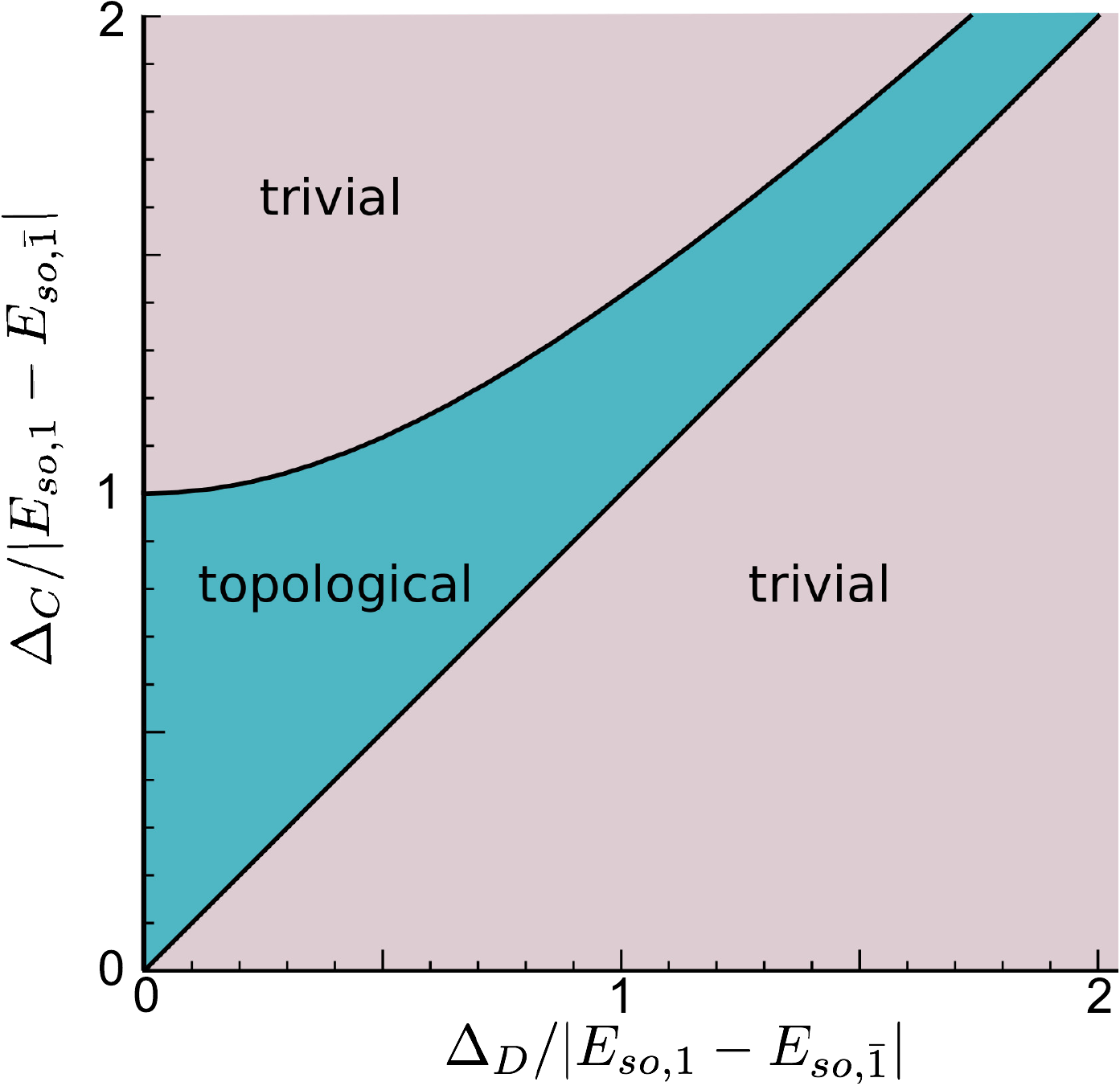}
	\caption{Topological phase diagram as a function of direct ($\Delta_D$) and crossed ($\Delta_C$) Andreev superconducting pairing amplitude. Topological phase transitions occur for $\Delta_D=\Delta_C$ and $\Delta_C^2=\Delta_D^2+(E_{so,1}-E_{so,\bar{1}})^2$ (black lines). The topological phase (blue area) hosts a Kramers pair of edge states, whereas in the trivial phase there are no edge states. The larger the difference of the spin-orbit energies of the two layers, the larger is the topological region.}
	\label{PhaseDiagram}
\end{figure}

The total Hamiltonian is given by $H=H_{1}+H_{\bar{1}}+H_D+H_C$ and can be rewritten in terms of Pauli matrices $H= \frac{1}{2}\int d^2 {\bf k}\ \Psi^{\dagger}_{{\bf k}} h({\bf k}) \Psi_{{\bf k}}$ with
\begin{align}
&h({\bf k}) = \epsilon_k \eta_3 +(\alpha_{+} + \alpha_- \tau_3)(\sigma_1 k_y - \eta_3 \sigma_2 k_x) \nonumber \\
&\hspace{35pt}- \Delta_D \eta_2 \sigma_2 - \Delta_C \tau_1 \eta_2 \sigma_2, \label{BdGHamiltonian}
\end{align}
where we have introduced $\alpha_{\pm}=(\alpha_1 \pm \alpha_{\bar{1}})/2$ and the Pauli matrices $\tau_i$, $\eta_i$, and $\sigma_i$ acting in layer, particle-hole, and spin space, respectively. One can check that $H$ is time-reversal invariant with the time-reversal operator given by $\Theta=- i \sigma_2 \mathcal{K}$, where $\mathcal{K}$ is the complex conjugation operator. The particle-hole symmetry operator is given by $\mathcal{P}=\eta_1$. Therefore, $H$ belongs to the DIII symmetry class which has a $\mathbb{Z}_2$ classification for 2D systems\cite{RyuSchnyder2010}.
The bulk spectrum of the bilayer setup is given by
\begin{align}
E^2_{\pm,\pm}(k) = &\tilde{\epsilon}_{\pm,k}^2 + \alpha_-^2 k^2 + \Delta_D^2+\Delta_C^2 \nonumber \\
& \pm 2\sqrt{\alpha_-^2 k^2 (\tilde{\epsilon}^2_{\pm,k}+\Delta_C^2)+\Delta_C^2 \Delta_D^2},
\end{align}
where  $\tilde{\epsilon}_{\pm,k}= \epsilon_k \pm\alpha_{+}k$. The bulk spectrum is gapped except in two special cases. First, if $\Delta_C=\Delta_D$, the bulk gap closes at $k=0$. Second, if $\Delta_C=\tilde{\Delta}_C$, where $\tilde{\Delta}_C^2= \Delta_D^2 + (E_{so,1}-E_{so,\bar{1}})^2$, the bulk gap closes at $k=2m \alpha_{+}/\hbar^2$. Here the bulk gap is not closed at one point in momentum space, as in the first case, but along a circular nodal line. Both bulk gap closings correspond to a topological phase transition. The topological phase diagram is presented in Fig. \ref{PhaseDiagram}, and contains one topological and two trivial regions. It is obvious to see that the system is in a trivial phase when $\Delta_C=0$ and $\Delta_D>0$. The system is therefore in the trivial phase for the parameter regime $\Delta_C<\Delta_D$. In the regime $\Delta_D<\Delta_C<\tilde{\Delta}_C$, the system is in the topological phase, where, as we will show below, a Kramers pair of edge states exists at each edge. For large $\Delta_C$, {\it i.e.} $\Delta_C>\tilde{\Delta}_C$, the edge states disappear, which is again a trivial phase.

\begin{figure}
	\centering
\includegraphics[width=\columnwidth]{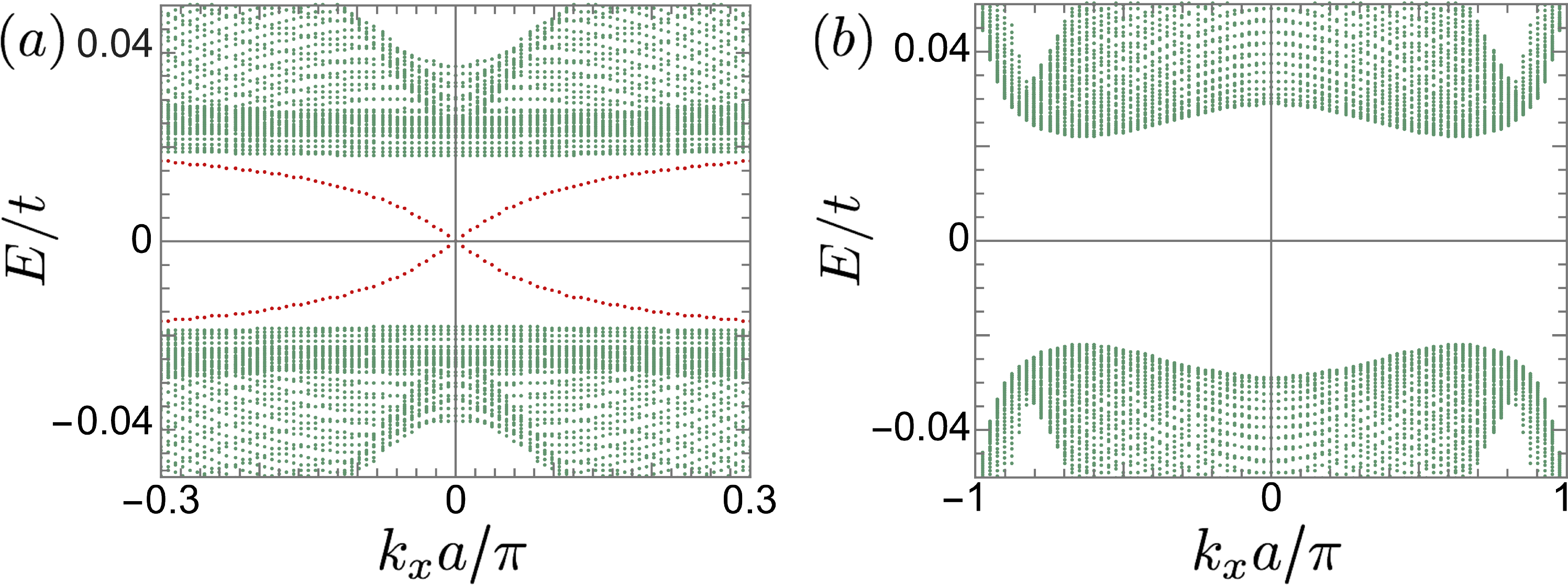}
	\caption{Energy spectrum of the bilayer setup in (a) the topological phase ($\Delta_C/t=0.11$) and (b) the trivial phase ($\Delta_C/t=0.22$). In both cases the bulk states (green) have a spectral gap, while a Kramers pair of edge states (red) is only present in (a). The edge states are localized at both edges and have a linear dispersion around $k_x=0$. The numerical parameters are chosen as $N_y=300$, $\mu/t=-4$, $\alpha_1/t=0.55$, $\alpha_{\bar{1}}/t=0.35$, and $\Delta_D/t=0.06$.}
	\label{SpectrumTRIPhase}
\end{figure}

\section{Helical Topological Superconductor} \label{HTSC}
\subsection{Helical edge states}

In order to confirm the phase diagram obtained from the bulk spectrum in the previous section, we now investigate a finite-size system and focus on the properties of the edges. We first solve the problem numerically by implementing a tight-binding model for the bilayer setup \cite{numerics}. Without loss of generality, the layers are taken to be finite along the $y$ direction, of length $L$ ($N_y$ lattice sites separated by lattice constant $a$), and translationally invariant along the $x$ direction, allowing us to use $k_x$ as a good quantum number. The Hamiltonian for this setup is given by $H=H_1+H_{\bar{1}}+H_D+H_C$ with
\begin{widetext}
	\begin{align}\label{HamiltonianTB}
 H_{\tau} &= \frac{1}{2}\sum_{k_x,n} \Big\{ \sum_{\sigma} \Big(-t c^{\dagger}_{k_x \tau (n+1) \sigma} c_{k_x\tau n \sigma} + t  c_{-k_x\tau n \sigma}  c^{\dagger}_{-k_x \tau (n+1) \sigma} + [-t \cos(k_x a_x) +\mu_{\tau}/2] c^{\dagger}_{k_x \tau n \sigma} c_{k_x\tau n \sigma} \nonumber \\
	& - [-t \cos(k_x a_x) +\mu_{\tau}/2] c_{-k_x \tau n \sigma} c^{\dagger}_{-k_x\tau n \sigma}+ \text{H.c.} \Big) 
	 + \tilde{\alpha}_{\tau}\Big[ i(c^{\dagger}_{k_x \tau (n+1) \uparrow}c_{k_x\tau n \downarrow} - c^{\dagger}_{k_x \tau (n-1) \uparrow} c_{k_x\tau n \downarrow})  + 2i \sin(k_x a_x) c^{\dagger}_{k_x \tau n \uparrow} c_{k_x\tau n \downarrow} \nonumber \\
	& i(c_{-k_x\tau n \downarrow}c^{\dagger}_{-k_x \tau (n+1) \uparrow} -  c_{-k_x\tau n \downarrow} c^{\dagger}_{-k_x \tau (n-1) \uparrow})  - 2i\sin(k_x a_x)  c_{-k_x\tau n \downarrow} c^{\dagger}_{-k_x \tau n \uparrow}+ \text{H.c.} \Big] \Big\}, \nonumber \\
	H_{C} &= \frac{\Delta_C}{2} \sum_{k_x} \sum_{n,\tau} \Big( c^{\dagger}_{k_x \tau n \sigma} [i \sigma_2]_{\sigma \sigma'} c^{\dagger}_{-k_x \bar{\tau} n \sigma'} + \text{H.c.}\Big), \nonumber \\
	H_D &= \frac{\Delta_D}{2} \sum_{k_x} \sum_{n,\tau} \Big( c^{\dagger}_{k_x \tau n \sigma} [i \sigma_2]_{\sigma \sigma'} c^{\dagger}_{-k_x \tau n \sigma'} + \text{H.c.}\Big).
	\end{align}
\end{widetext}

Here, $t$ denotes the hopping amplitude. The operator $c_{k_x \tau n \sigma}$ acts on an electron at position $y=n a$ in the layer $\tau(=-\bar{\tau})$ with momentum $k_x$ and spin projection $\sigma$. The spin-flip hopping amplitude is related to the SOI parameter by $\tilde{\alpha}=\alpha/2 a_y$ \cite{Diego2013}.
Solving the Hamiltonian $H$, one finds that there are no edge states in the regimes $\Delta_C<\Delta_D$ and $\Delta_C>\tilde{\Delta}_C$ [see Fig. \ref{SpectrumTRIPhase}(b)], which confirms that these parameter regimes correspond to the trivial phases (see Fig. \ref{PhaseDiagram}). In the regime $\Delta_D<\Delta_C<\tilde{\Delta}_C$ both edges host a Kramers pair of subgap states. These states are localized on the edges and have a linear dispersion around $k_x=0$ [see Fig. \ref{SpectrumTRIPhase}(a)]. This is the hallmark of a helical TSC and confirms our expectation that this parameter regime corresponds to the topological phase.

\begin{figure}[t]
	\centering
	\includegraphics[width=.85\columnwidth]{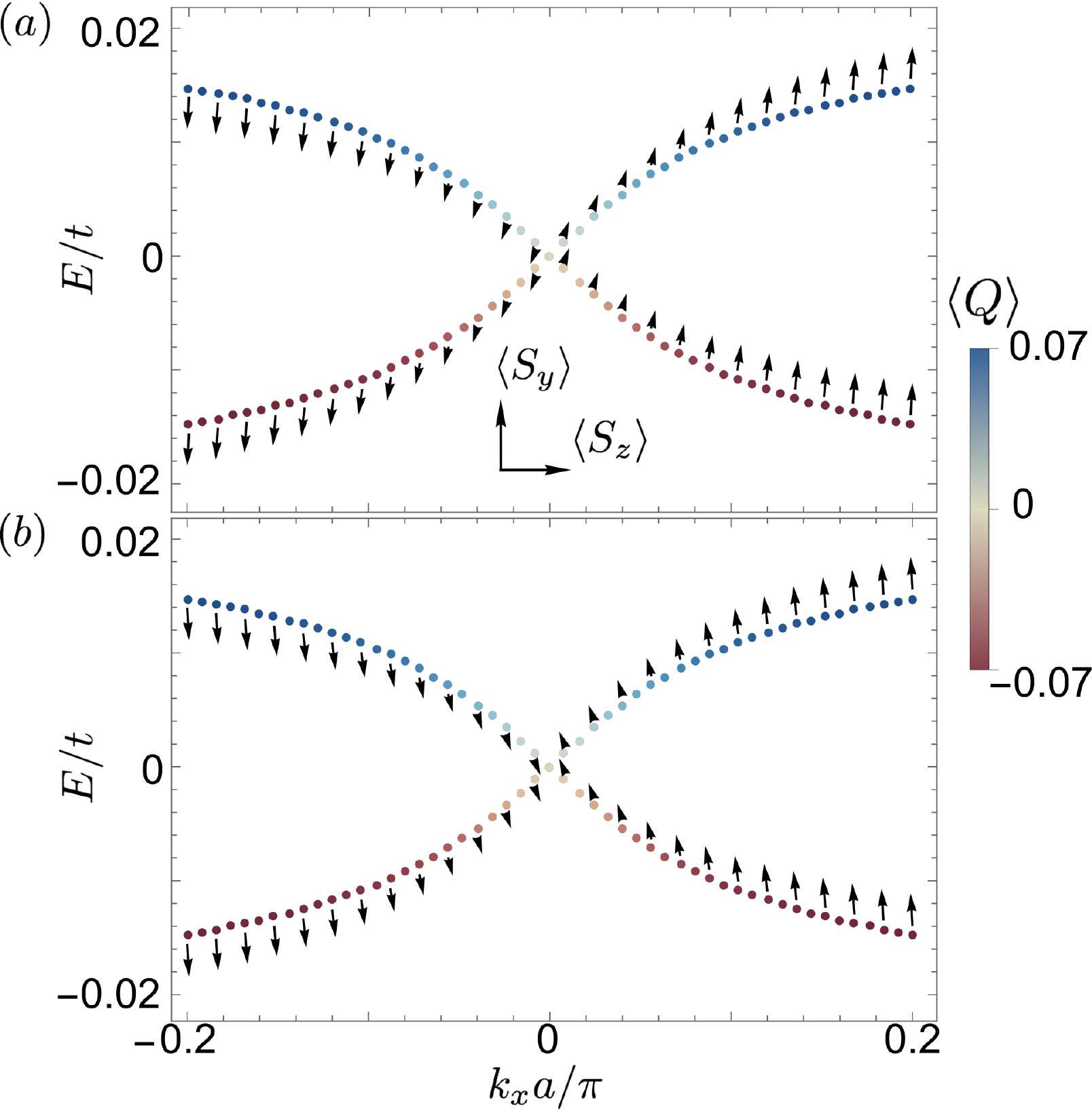}
	\caption{Energy spectrum of helical edge states (a) on the left edge ($y=0$) and (b) on the right edge ($y=L$). The average of the charge operator is encoded in the coloring of the data points and given in units of $e$. One can see that for a given energy $\braket{Q(k_x)} = \braket{Q(-k_x)}$ as is expected, since time-reversal does not invert the sign of the charge. The average spin as a function of $k_x$ is encoded in the black arrows. The average spin component along the $x$ axis is zero for all momenta, while the other components are non-zero, and for the Kramers partner the relation $\braket{S_i(k_x)}_1= - \braket{S_i(-k_x)}_2$ holds. The wavefunctions on opposite edges are related by the reflection symmetry operator $\mathcal{I}=\sigma_y$. Thus, states on opposite edges, which propagate in the same direction, have the same average spin along the $y$ axis whereas their average spins along the $z$ axis are opposite. The numerical parameters are the same as in Fig. \ref{SpectrumTRIPhase}.}
	\label{ExpectationValueCharge}
\end{figure}

Further, we investigate the spin and charge properties of the edge states. Since in our setup spin and charge are not conserved quantities, we calculate the expectation value of the spin (charge) operator ${\bf S}$ ($Q$). In the following, $\braket{O}_{\beta} = \braket{\Phi_{\beta}|O|\Phi_{\beta}}$ denotes the expectation value for some operator $O$, where the ket $\ket{\Phi_{\beta}}$ describes the two edge states labeled by $\beta \in \{1,2\}$ (for further details we refer to Appendix A). At $k_x=0$, where the system can be mapped to the 1D analogue of our setup \cite{KlinovajaPF2014,Schrade2017}, the edge states have zero average spin and charge. This is consistent with previous works. For all other values of $k_x$ the expectation values are generally non-zero and, as is expected, for fixed energy the two Kramers partners have the same charge. The average of the $x$ component of the spin vanishes, $\braket{S_x}=0$, for all values of $k_x$, while the remaining components satisfy $\braket{S_i(k_x)}_1= - \braket{S_i(-k_x)}_2$ (see Fig. \ref{ExpectationValueCharge}). In this sense, the edge states are helical and protected from back-scattering by time-reversal symmetry. Note that the edge states on the left and the right edge are connected by reflection symmetry, where the symmetry operator is given by $\mathcal{I}=\sigma_y$. Thus, the wavefunctions of the edge states on opposite edges and their properties are related by $\mathcal{I}$. Therefore, states on opposite edges that propagate in the same direction have the same average spin projection on the $y$ axis. In contrast, their average spin projections on the $z$ axis are opposite (differ by a minus sign), see Fig. \ref{ExpectationValueCharge}.

\begin{figure}[t]
	\centering
	\includegraphics[width=.7\columnwidth]{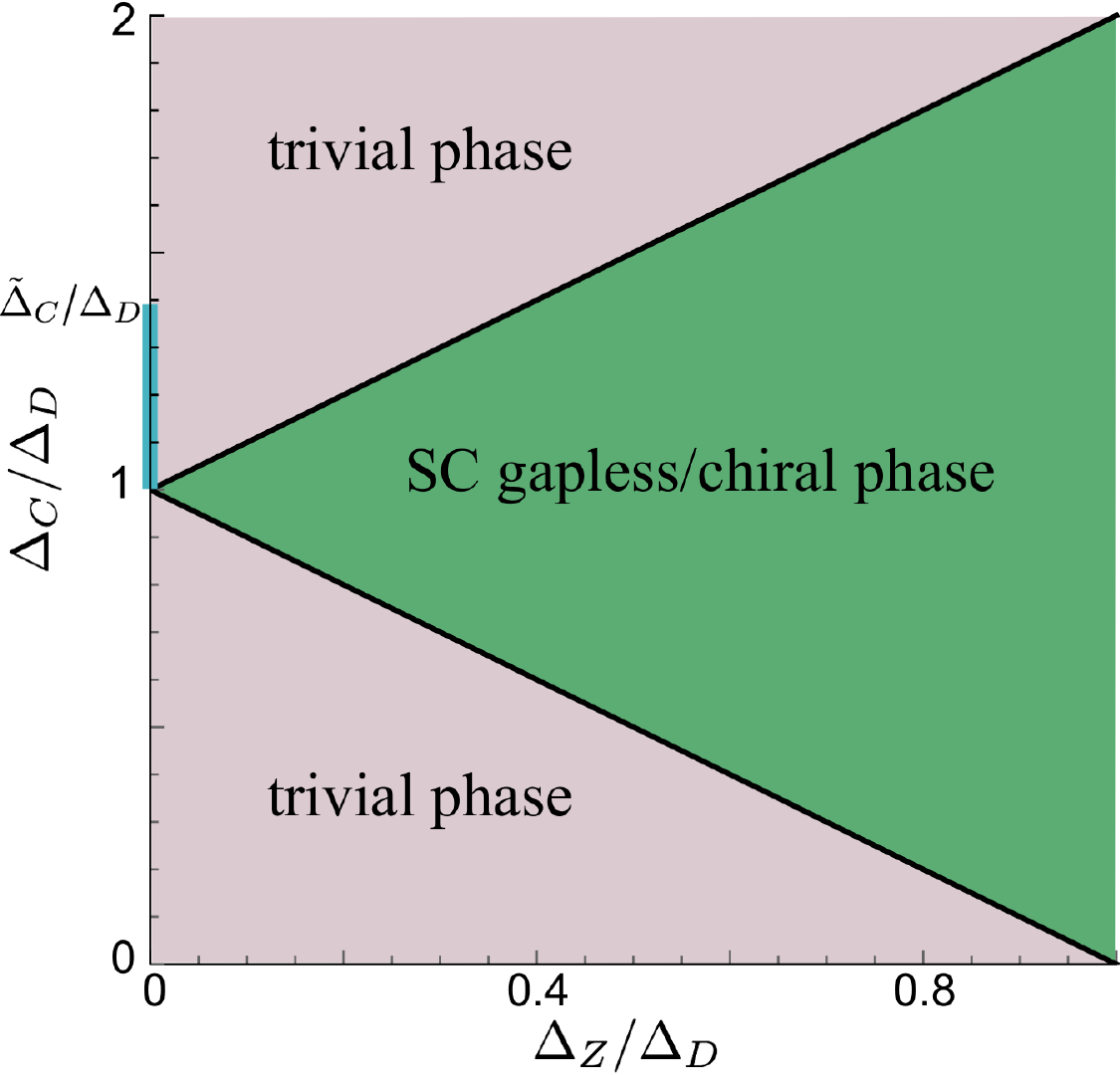}
	\caption{Topological phase diagram as a function of the Zeeman energy ($\Delta_Z$) and the crossed Andreev superconducting pairing amplitude ($\Delta_C$). Topological phase transitions occur for $\Delta_C = |\Delta_D \pm \Delta_Z|$ (black lines) and for $\Delta_C=\tilde{\Delta}_C$ and $\Delta_Z=0$. In the absence of a Zeeman field there exists a helical topological phase for $\Delta_D<\Delta_C<\tilde{\Delta}_C$, which is indicated by the blue line.  We note that $\tilde{\Delta}_C$ depends on both $\Delta_D$ and the difference in SOI energies, and in this plot we assume $\Delta_D=|E_{so,1}-E_{so,\bar{1}}|$ such that $\tilde{\Delta}_C/\Delta_D=\sqrt{2}$. In the green region the system is in a 2D superconducting gapless (chiral) phase for an in-plane (out-of-plane) Zeeman field. For details on the transition from the gapless to the chiral phase as a function of the out-of-plane angle we refer to the main text.}
	\label{PhaseDiagZeeman}
\end{figure}

\subsection{Effective low-energy Hamiltonian}
As shown in the previous section, the topological phase hosts a Kramers pair of helical edge states with a linear dispersion around $k_x=0$ on both edges. Next, we derive the effective low-energy Hamiltonian describing the properties of these states localized at the left ($y=0$) edge. As in the tight-binding model, we assume that the system is translationally invariant along the $x$ direction and we solve an effectively 1D Hamiltonian parametrized by $k_x$. For the moment, we assume that the right edge ($y=L$) is infinitely far away. This assumption allows us to treat the edge states as if the width of the sample was much larger than the localization length $\xi$ of the edge states. Due to particle-hole symmetry, the edge states are at zero energy at $k_x=0$ [see Fig. \ref{SpectrumTRIPhase}(a)], and we first determine the wavefunctions at this special point. In a next step, we treat the $k_x$-terms  perturbatively for $k_x \xi \ll 1$, and keep only terms linear in $k_x$. Using $h(0,k_y)$ [see Eq. \eqref{BdGHamiltonian}] we obtain two wavefunctions $\Phi_{\beta}$ ($\beta \in \{1,2\}$) in the strong SOI regime for $\alpha_1 \gg \alpha_{\bar{1}}$ (see Appendix B). These two wavefunctions correspond to a Kramers pair of edge states and  exist only if $\Delta_D<\Delta_C<\tilde{\Delta}_C$, which is consistent with our previous results. The wavefunctions have support in momentum space around $k_{F \tau} = 2 k_{so \tau}$ and $k_{F,i}=0$. The corresponding localization length $\xi$ is determined by the bulk gaps at these Fermi points\cite{Klinovaja2012} and, thus, is given by the maximum of four length scales: $\xi_{\tau}= \hbar v_{F,\tau}/\Delta_D$ and
\begin{equation}
\xi_{\pm} = \frac{2 \hbar v_{F,1} v_{F,\bar{1}}}{ \sqrt{\Delta_D^2 (v_{F,1}-v_{F,\bar{1}})^2 + 4 \Delta_C^2 v_{F,1} v_{F,\bar{1}}}\pm \Delta_D(v_{F,1}+v_{F,\bar{1}}) },
\end{equation}
respectively, where we have introduced the Fermi velocities $v_{F \tau}= \alpha_{\tau}/\hbar $.
\begin{figure}[!b]
	\centering
	\includegraphics[width=\columnwidth]{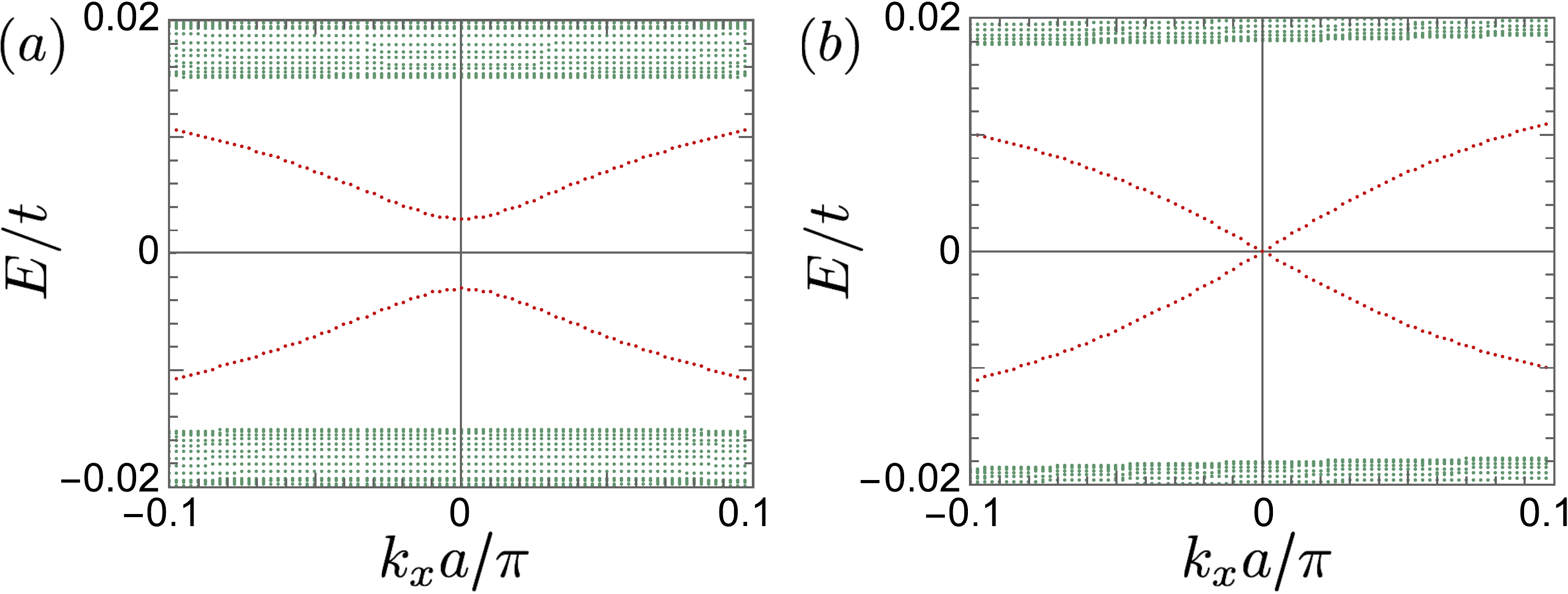}
	\caption{Energy spectrum in the presence of an in-plane Zeeman field aligned (a) in the $x$ direction ($\phi=0$) and (b) in the $y$ direction ($\phi=\pi/2$). Green (red) dots represent bulk (edge) states. The edge states become gapped once the Zeeman field has a component along the respective edge. The size of the gap opened in the edge state spectrum is given by the projection of the Zeeman field on the propagation direction, $\Delta E= \Delta_Z\cos\phi$. The numerical parameters are chosen as $N_x=800$, $\alpha_1/t=0.35$, $\alpha_{\bar{1}}/t=0.15$, $\mu=-4t$, $\Delta_D/t=0.06$, $\Delta_C/t=0.12$, and $\Delta_Z/t=0.006$}
	\label{SpectraMagneticField}
\end{figure}

\begin{figure*}[!t]
	\centering
\includegraphics[width=0.9\textwidth]{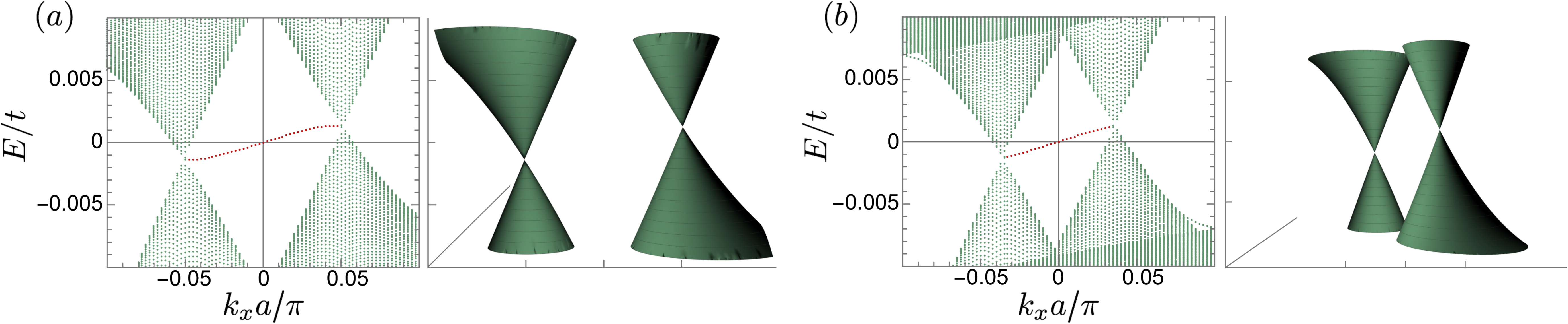}
	\caption{Energy spectra in the presence of an inplane Zeeman field (a) $\phi=\pi/2$ and (b)  $\phi=\pi/4$. In the left subfigures the bulk (edge) states are represented by green (red) dots and obtained from the tight-binding model. In the right subfigure, we show the bulk spectrum of the translationally-invariant system. The line connecting centers of two Weyl cones is orthogonal to the direction of the Zeeman field. The degenerate edge states (one per edge) connect the two Weyl cones. The numerical parameters are chosen as $N_y=2000$, $\alpha_1/t=0.35$, $\alpha_{\bar{1}}/t=0.15$, $\mu=-4t$, $\Delta_D/t=0.05$, $\Delta_C=\Delta_D$, and $\Delta_Z/t=0.025$.}
	\label{InplaneAngles}
\end{figure*}

For $k_x \xi\ll 1 $, the perturbation term linear in $k_x$ is given by $h_{k_x}=(\alpha_{+} + \alpha_- \tau_3) \eta_3 \sigma_2 k_x$ [see Eq. \eqref{BdGHamiltonian}], and the first order correction to the energy can be found by calculating the matrix elements $\braket{\Phi_{\beta}|h_{k_x}|\Phi_{\beta'}}$. As a result the effective low-energy Hamiltonian is given by
\begin{equation}\label{EffectiveHamiltonian}
h_{eff}= \hbar \tilde{v} \beta_3 k_x.
\end{equation}
Here $\beta_3$ is the third Pauli matrix acting in the low-energy subspace spanned by $\Phi_{\beta}$ and $\tilde{v}$ the effective Fermi velocity. The effective Hamiltonian $h_{eff}$ has a form typical for helical TSCs. The helical edge states have a linear dispersion inside the bulk gap.

\section{ Effect of a Zeeman field. 2D gapless vs. chiral superconducting topological phase} \label{MagneticField}

Having discussed the helical topological superconducting phase in the previous section, we now investigate the possible topological phases in the presence of an effective Zeeman field, which could arise due to presence of polarized magnetic impurities in both layers, similarly, as was already discussed in the literature for magnetic islands on superconductors, Weyl semimetals or quantum anomalous Hall effect \cite{Li2016,Menard2017,Chen2010,Hosseini2015,Chang2015_2,Nunner2008,Sacramento2012,Raymond2015}. We consider an effective Zeeman field in the direction determined by the unit vector ${\bf n}=(\cos \phi \cos \theta, \sin \phi \cos \theta,\sin \theta)^T$, then the Zeeman term reads
\begin{equation}\label{HamiltonianZeeman}
H_Z = \Delta_Z \sum_{\tau} \sum_{\sigma, \sigma'} \int d^2 {\bf k} \ \psi_{\tau \sigma, {\bf k}}^{\dagger} [\boldsymbol{n} \cdot \boldsymbol{\sigma} ]_{\sigma \sigma'} \psi_{\tau \sigma', {\bf k}},
\end{equation}
where $\Delta_Z$ denotes the Zeeman energy. 

Again, we first focus on the phase diagram in the presence of an in-plane Zeeman field ($\theta=0$), which breaks time-reversal symmetry, see Fig. \ref{PhaseDiagZeeman}. If $\Delta_Z>\Delta_D$, the superconductivity is strongly suppressed and the system is gapless.  If $\Delta_C=|\Delta_Z \pm \Delta_D|$, the Zeeman term leads to a closing of the bulk gap at ${\bf k}=0$ (black lines in Fig. \ref{PhaseDiagZeeman}).
If $\Delta_C<\Delta_D$ and $\Delta_Z=0$ the system is in a trivial phase as was shown above, and therefore, in the whole region $\Delta_C<|\Delta_Z-\Delta_D|$ of the phase diagram, the system is in the trivial phase. In the regime $\Delta_D<\Delta_C<\tilde{\Delta}_C$ and $\Delta_Z=0$ the system is in the helical TSC phase with a Kramers pair of gapless edge states (blue line in Fig. \ref{PhaseDiagZeeman}). When $\Delta_Z>0$, time-reversal symmetry is broken and, thus, the helical edge states are no longer protected against backscattering. If the Zeeman field has a component along a given edge, it leads to a coupling between the Kramers partners and thereby gaps out the edge states. The size of the gap is given by the projection of the Zeeman field on the given edge, which for our tight-binding model translates into $\Delta E= \Delta_Z \cos \phi$ (see Fig. \ref{SpectraMagneticField}). In the special case of $\phi=\pi/2$, the pair of edge states propagating along the $x$ axis stay gapless. However, if the system was finite in both $x$ and $y$ directions, one can make use of the spatial symmetries of the setup to conclude that the spectrum of the edge states propagating along the $y$ direction is gapped. Thus, we conclude that a weak in-plane Zeeman field leads to a gap in the spectrum of the helical edge states at least along one of the edges, so the region $\Delta_C>|\Delta_Z+\Delta_D|$ corresponds to a trivial phase. 
\begin{figure*}[t!]
	\centering
		\includegraphics[width=0.9\textwidth]{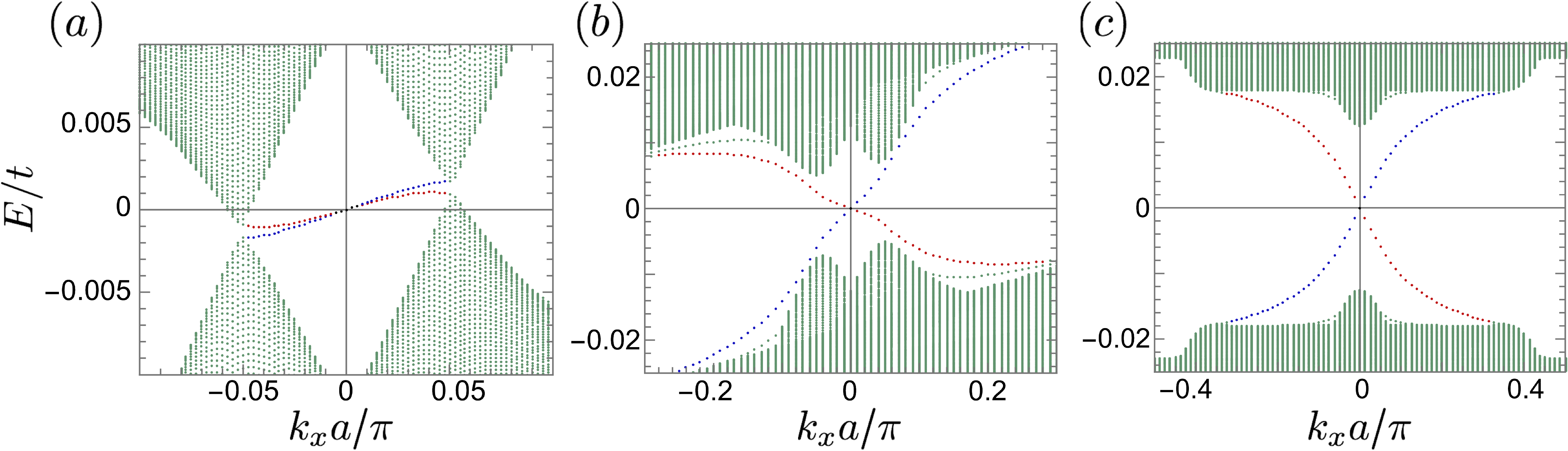}
	\caption{Energy spectra in the presence of an out-of-plane Zeeman field for $\phi=\pi/2$ and (a) $\theta=\pi/90$, (b) $\theta=\pi/6$, and (c) $\theta=\pi/2$. The Zeeman field component along the $z$ direction leads to an opening of a gap in the bulk spectrum (green dots) at the position of the nodes. The gap increases with the out-of-plane angle $\theta$. At the same time the degeneracy of the edge states (see Fig. \ref{InplaneAngles}) gets lifted as the out-of-plane angle increases, and the edge states become chiral, i.e. counterpropagating on opposite edges (red and blue dots) for $\theta>\pi/15$. The other numerical parameters are the same as in Fig. \ref{InplaneAngles}.}
	\label{SpecChiral}
\end{figure*}

Next, we analyze the parameter region of the phase diagram that is bounded by the two gap closing lines $\Delta_C=|\Delta_Z \pm \Delta_D|$ (green area in Fig. \ref{PhaseDiagZeeman}). In this regime, the bulk spectrum contains two Weyl cones. Their position in the Brillouin zone is determined by the polar angle $\phi$ of the Zeeman field: the line connecting the two Weyl cones is orthogonal to the direction of the effective Zeeman field (see insets in Fig. \ref{InplaneAngles}). Localized edge states connect the two Weyl nodes in momentum space, where they coexist with bulk modes. Since the nodes are not at zero energy the edge states have a finite group velocity (see Fig. \ref{InplaneAngles}). There is one state per edge and they have the same direction of propagation on opposite edges, {\it i.e.} the system hosts unidirectional edge states \cite{Wong2013,Daido2017}. As is the case for Weyl semimetals, in order to determine whether edge states appear on a given edge, one has to project the Weyl nodes onto the edge direction; edge states appear only on edges where the Weyl nodes are not projected onto the same point. For our tight-binding model, this means that the unidirectional edge states do not appear on the edge along the $x$ direction if $\phi=0$. A gap can be opened in the 2D Weyl spectrum if a perturbation proportional to $\sigma_z$ is added, {\it i.e.} an effective Zeeman field has an out-of-plane component along the $z$ direction. Thus, it is interesting to investigate the evolution of the edge states as the out-of-plane angle $\theta$ of the Zeeman field is increased. The larger the angle the larger is the gap, which is opened in the spectrum of Weyl cones. For small $\theta$, the bulk spectrum stays gapless and the edge states remain unidirectional although their two-fold degeneracy gets lifted (see Fig. \ref{SpecChiral}). For the parameter settings used in our numerics this holds for $\theta \leq \pi/15$. If $\theta$ is larger than this threshold, the bulk spectrum is fully gapped and the edge states evolve from unidirectional to chiral edge states (see Fig. \ref{SpecChiral}).
We conclude that a 2D gapless topological phase with unidirectional edge states is only achieved if the out-of-plane angle of the Zeeman field is 'small', otherwise the system is a fully gapped chiral topological superconductor.

\section{Conclusions}
We have shown how three distinct topological superconducting phases can be engineered in a heterostructure composed of two 2DEG layers with strong Rashba SOI and an $s$-wave superconductor sandwiched between these two layers. Such a setup could be realized in semiconductor superlattices\cite{Deutschmann2001,Nitta1997,Dettwiler2017} or using (quasi-) 2D materials with strong SOI\cite{Novoselov2016,Sachs2013,Novoselov2012,Avsar2015,Fulop2017}. 
 
In the time-reversal symmetric case, a pair of helical edge states exists when crossed Andreev superconducting pairing is dominant. When time-reversal symmetry is broken by a Zeeman field, the system can potentially be in two distinct topological phases. If the field is in-plane or if the out-of-plane component is small enough, a gapless topological superconducting phase with unidirectional edge states can be realized. If the out-of-plane angle is large enough, the system can be tuned into a gapped topological superconducting phase with chiral edge states.
It is well-known that the TSCs with a bulk gap are the most stable against disorder. The helical TSC is, due to Kramers theorem, stable against non-magnetic disorder, while a chiral TSC is stable against both non-magnetic and magnetic disorder. However, in both cases, this only holds as long as the disorder strength is not comparable to the size of the superconducting gap and Weyl cones are well separated, which is the case for long-range disorder. \cite{Burkov2011} 

\section*{Acknowledgments}
We acknowledge support from the Swiss National Science Foundation and NCCR QSIT, and  the Marie Sklodowska-Curie Innovative Training Network (ITN-ETN) Spin-NANO. This project has received funding from the European Union’s Horizon 2020 research and innovation program (ERC Starting Grant, grant agreement  No 757725). We acknowledge helpful discussions with Christopher Reeg and Denis Chevallier.

\begin{appendix}
\section{Numerical evaluation of charge and spin expection values}
In the main text we discussed the expectation values of the spin (charge) operator for the helical edge states. In this paragraph we complete the discussion by providing the detailed expressions. Solving numerically the tight-binding model defined by Eq. \eqref{HamiltonianTB}, we obtain $8N_y$ energy states for each fixed momentum $k_x$, i.e. the energy states can be parametrized as $E_n(k_x)$ with the corresponding eigenfunctions $\Phi_{n}(j,k_x)$, where $n \in \{1,\dots,8N_y\}$ is the band index and $j\in \{1,\dots,N_y\}$ denotes the lattice site. Here, $\Phi_n$ is an eight component spinor in Nambu space written in the basis $(c_{k_x 1 \uparrow n},c_{k_x 1 \downarrow n},c^{\dagger}_{-k_x 1 \uparrow n},c^{\dagger}_{-k_x 1 \downarrow n},c_{k_x \bar{1} \uparrow n},c_{k_x \bar{1} \downarrow n},c^{\dagger}_{-k_x \bar{1} \uparrow n},c^{\dagger}_{-k_x \bar{1} \downarrow n})$.

The average spin and charge of the $n$-th energy eigenstate with $k_x$ fixed is computed as\cite{Szumniak2017}
\begin{align}
\braket{{\bf S}(k_x)}_n &= \sum_{j} \Phi^{\dagger}_n(j,k_x) {\bf S} \Phi_{n}(j,k_x), \nonumber \\
\braket{Q(k_x)}_n &= \sum_{j} \Phi^{\dagger}_n(j,k_x) Q \Phi_{n}(j,k_x),
\end{align}
where ${\bf S}$ ($Q$) is the spin (charge) operator measured in units of $\hbar/2$ ($e$) and represented in Nambu space $S_i= \text{diag}(\sigma_i,-\sigma_i^T,\sigma_i,-\sigma_i^T)$ [$Q=\text{diag}(\sigma_0,-\sigma_0,\sigma_0,-\sigma_0)$], where $\sigma_i$ are the Pauli matrices and $\sigma_0$ is the $2 \times 2$ identity matrix.

\section{Derivation of the effective low-energy Hamiltonian}
In this section we present the detailed calculation of the wavefunctions at $k_x=0$ and derive the effective low-energy Hamiltonian describing helical edge states (see Eq. \eqref{EffectiveHamiltonian} in Sec. \ref{HTSC}).

The wavefunctions at $k_x=0$ should be related by time-reversal symmetry $\Theta$ and also can be chosen to respect the particle-hole symmetry \cite{KlinovajaPF2014}. As a result, they can be represented  in the basis $(\psi_{1 \uparrow},\psi_{1 \downarrow},\psi^{\dagger}_{1 \uparrow},\psi^{\dagger}_{1 \downarrow},\psi_{\bar{1} \uparrow},\psi_{\bar{1} \downarrow},\psi^{\dagger}_{\bar{1} \uparrow},\psi^{\dagger}_{\bar{1} \downarrow})$ in the following form \cite{Klinovaja2012}
\begin{equation}\label{MajoranaWFDefinition}
\Phi_{1}(y)= \left(\begin{array}{c} f_1(y) \\ g_1(y) \\ f_1^*(y) \\ g_1^*(y) \\ f_{\bar{1}}(y) \\ g_{\bar{1}}(y) \\ f_{\bar{1}}^*(y) \\ g_{\bar{1}}^*(y)\end{array}\right), \
\Phi_{2}(y)= \left(\begin{array}{c} g^*_1(y) \\-f^*_1(y) \\ g_1(y) \\ -f_1(y) \\ g^*_{\bar{1}}(y) \\ -f^*_{\bar{1}}(y) \\ g_{\bar{1}}(y) \\-f_{\bar{1}}(y)\end{array}\right).
\end{equation}

We start with $h(0,k_y)$ [see Eq.\eqref{BdGHamiltonian}], which after the unitary transformation $U=e^{-\frac{i \pi}{4} \sigma_2}$ reads
\begin{align}\label{H1D}
h(0,k_y) = &\frac{\hbar^2 k_y^2}{2m}  \eta_3 + (\alpha_{+} + \alpha_- \tau_3)\sigma_3 k_y - \Delta_D \eta_2 \sigma_2\nonumber \\
 &-\Delta_C \tau_1 \eta_2 \sigma_2.
\end{align}

In the next step, we linearize \cite{Linear} the spectrum around the Fermi points $k_i=0$, $k_{F \tau}= \pm 2 k_{so,\tau}$ such that the field operators $\psi_{\tau \sigma}$ are approximated as
\begin{align}
\psi_{\tau \uparrow} &= L_{\tau \uparrow}(y) + R_{\tau \uparrow}(y) e^{i k_{F \tau}y}, \nonumber \\
\psi_{\tau \downarrow} &= L_{\tau \downarrow}(y)e^{-i k_{F \tau}y} + R_{\tau \downarrow}(y),
\end{align}
where $R_{\tau \sigma}(y)$ [$L_{\tau \sigma}(y)$] are slowly varying right (left) moving fields with spin projection $\sigma$ along the $x$-axis (due to the unitary transformation $U$ above) in the $\tau$ layer. In the limit of strong SOI energy and $\alpha_1 \gg \alpha_{\bar{1}}$, the crossed Andreev pairing term only couples fields at $k=0$ \cite{KlinovajaPF2014}. As a result, the linearized Hamiltonian reads
\begin{align}
\bar{h} =  &v_{F 1} \hat{k} \frac{1+\tau_3}{2} \rho_3 + v_{F \bar{1}} \hat{k} \frac{1-\tau_3}{2}\rho_3 - \Delta_D \eta_2 \sigma_2 \rho_1 \nonumber \\
 &- \Delta_C \tau_1 \eta_2 (\sigma_2 \rho_1 - \sigma_1 \rho_2)/2,
\end{align}
with $\hat{k}=-i \hbar \partial_y$ the momentum operator around the Fermi points and $\rho_i$ acting in left/right mover space. We make the Ansatz $\phi_{\xi}(y)= \phi_{\xi} e^{-y/\xi}$ with $\phi_{\xi}$ being a 16-component vector and search for decaying zero energy solutions satisfying $\bar{h} \phi_{\xi}=0$. We find eight decaying eigenmodes with localization lengths given by
\begin{align}
\xi_{1} &= \frac{\hbar v_{F,1}}{\Delta_D}, \ \xi_{2} =  \frac{\hbar v_{F,\bar{1}}}{\Delta_D}, \\
\xi_{3} &= \frac{2 \hbar v_{F,1} v_{F,\bar{1}}}{-\Delta_D(v_{F,1}+v_{F,\bar{1}}) + \sqrt{\Delta_D^2 (v_{F,1}-v_{F,\bar{1}})^2 + 4 \Delta_C^2 v_{F,1} v_{F,\bar{1}}}}, \nonumber \\
\xi_{4} &= \frac{2 \hbar v_{F,1} v_{F,\bar{1}}}{\Delta_D (v_{F,1}+v_{F,\bar{1}}) + \sqrt{\Delta_D^2 (v_{F,1}-v_{F,\bar{1}})^2 + 4\Delta_C^2 v_{F,1} v_{F,\bar{1}}}}.\nonumber 
\end{align}
By imposing vanishing boundary condition $\Phi_\beta(0) \overset{!}{=} 0$ on a linear combination of these eigenmodes,  we find that the edge state wavefunctions are determined by 
\begin{align}
f_1(y)&=i g_1^*(y)= -i (e^{-y/\xi_4}-e^{-ik_{F1}y}e^{-y/\xi_1}), \nonumber \\
f_{\bar{1}}(y) &= i g_{\bar{1}}^*(y)= -\frac{i}{g_-}(e^{-ik_{F \bar{1}}y} e^{-y/\xi_2}-e^{-y/\xi_4}), \label{Functions}
\end{align}
with 
\begin{equation}
g_-=\frac{\Delta_D(v_{F1}-v_{F \bar{1}})+\sqrt{\Delta_D^2(v_{F1}-v_{F \bar{1}})^2 +4 \Delta_C^2 v_{F 1} v_{F \bar{1}}}}{2 \Delta_C v_{F1}}.
\end{equation}
In the limit $v_{F1}=v_{F \bar{1}}$ this solution reproduces the wavefucntions of Kramers pair of Majorana fermions found in Ref. \onlinecite{KlinovajaPF2014}. We note that if $\Delta_D>\Delta_C$, eight eigenmodes are linearly independent and their linear combination cannot satisfy vanishing boundary condition. Therefore, edge states do not exist in this regime.

For $k_x \xi \ll 1$, the perturbation linear in $k_x$ reads $h_{k_x}=(\alpha_{+} + \alpha_- \tau_3)\eta_3 \sigma_2 k_x/2 $. The first order correction to the energy is calculated by evaluating the matrix elements $\braket{\Phi_{n}|h_{k_x}|\Phi_{n'}}$. Using Eqs. \eqref{MajoranaWFDefinition} and \eqref{Functions}, we arrive at
	\begin{align}
	\braket{\Phi_{1}|h_{k_x}|\Phi_{1}} &= -2 k_x \sum_{\tau=\{1,\bar{1}\}} \alpha_{\tau}\int_0^{\infty} dy \ \big([f_{\tau}^*]^2 +[f_{\tau}]^2\big) \nonumber \\
	&= - \braket{\Phi_{2}|h_{k_x}|\Phi_{2}}, \nonumber \\
	\braket{\Phi_{1}|h_{k_x}|\Phi_{2}} &= -2i k_x \sum_{\tau=\{1,\bar{1}\}} \alpha_{\tau}\int_0^{\infty} dy \ \big( [f_{\tau}^*]^2-[f_{\tau}]^2\big) \nonumber \\
	&= \braket{\Phi_{2}|h_{k_x}|\Phi_{1}},
	\end{align}
where for compactness we express the matrix elements in terms of $f_1$($f_{\bar{1}}$) only. Decomposing the functions $f_{\tau}(y)=\text{Re}[f_{\tau}(y)] + i \text{Im}[f_{\tau}(y)]$ into real and imaginary parts we rewrite the matrix elements above as
\begin{align}
\braket{\Phi_{1}|h_{k_x}|\Phi_{1}} &=-4  k_x \sum_{\tau=\{1,\bar{1}\}} \alpha_{\tau}\int_0^{\infty} dy \ \big([\text{Re}(f_{\tau})]^2 - [\text{Im}(f_{\tau})]^2\big) , \nonumber\\
\braket{\Phi_{1}|h_{k_x}|\Phi_{2}} &=-8  k_x \sum_{\tau=\{1,\bar{1}\}} \alpha_{\tau}\int_0^{\infty} dy \ \text{Re}({f_{\tau}}) \text{Im}({f_{\tau}}) , \label{MatrixElements}
\end{align}
which now can be seen to be purely real. For completeness we give the expressions for the decomposition of $f_{\tau}$ into real and imaginary parts
\begin{align}
\text{Re}(f_1) &= \sin(k_{F1}y) e^{-y/\xi_1}, \nonumber \\
\text{Im}(f_1) &= \cos(k_{F1}y) e^{-y/\xi_1} - e^{-y/\xi_4}, \nonumber \\
\text{Re}(f_{\bar{1}}) &= -\frac{\sin(k_{F \bar{1}}y) e^{-y/\xi_2}}{g_-}, \nonumber\\
\text{Re}(f_{\bar{1}}) &= \frac{1}{g_-} \big[e^{-y/ \xi_4} - \cos(k_{F \bar{1}}y)e^{-y/\xi_2}\big].
\end{align}

Plugging these equations into \eqref{MatrixElements}, we find the following effective Hamiltonian
\begin{equation}
h_{\text{eff}} = \left( \begin{array}{cc} \braket{\Phi_1|h_{k_x}|\Phi_1} & \braket{\Phi_1|h_{k_x}|\Phi_2} \\ \braket{\Phi_2|h_{k_x}|\Phi_1} & \braket{\Phi_2|h_{k_x}|\Phi_2}\end{array} \right)=\left( \begin{array}{cc} A & B \\ B & -A\end{array} \right)k_x.
\end{equation}
where the real constants $A$ and $B$ are found to be given by
\begin{widetext}
	\begin{align}
	A  &= -4  \Big[\frac{2\alpha_1 \xi_1}{1+k_{F1}^2 \xi_1^2} + 2\alpha_1 \xi_4  - \frac{8 \alpha_1 \xi_1 \xi_4 (\xi_1+\xi_4)}{(\xi_1+\xi_4)^2+k_{F1}^2 \xi_1^2 \xi_4^2}+\frac{2\alpha_{\bar{1}}\xi_2}{g_-^2(1+k_{F \bar{1}}^2\xi_2^2)} +\frac{2 \alpha_{\bar{1}} \xi_4}{g_-^2}  - \frac{8 \alpha_{\bar{1}}\xi_2 \xi_4(\xi_2+\xi_4)}{g_-^2[(\xi_2+\xi_4)^2+k_{F \bar{1}}^2 \xi_2^2 \xi_4^2]}\Big], \nonumber \\
	B &= 2 \Big[\alpha_1 k_{F1}\Big(\frac{4 \xi_1^2 \xi_4^2}{(\xi_1+\xi_4)^2+k_{F1}^2 \xi_1^2 \xi_4^2}-\frac{\xi_1^2}{1+k_{F1}^2 \xi_1^2}\Big) + \frac{\alpha_{\bar{1}} k_{F_{\bar{1}}}}{g_-^2}\Big(\frac{4 \xi_2^2 \xi_4^2}{(\xi_2+\xi_4)^2+k_{F1}^2 \xi_2^2 \xi_4^2}-\frac{\xi_2^2}{1+k_{F1}^2 \xi_2^2}\Big)\Big].
	\end{align}
\end{widetext}
We find that this matrix is diagonal in the basis $\Phi^{\pm}$ with eigenvalues $\pm \hbar \tilde{v} k_x$, where $\tilde{v}=\sqrt{A^2+B^2}/\hbar$ and
\begin{equation}\label{EdgeModes}
\Phi^{\pm} = \frac{B \Phi_1 \pm (\sqrt{A^2+B^2} \mp A) \Phi_2}{\sqrt{B^2+(\sqrt{A^2+B^2} \mp A)^2}}.
\end{equation}
\end{appendix}
\pagebreak

\end{document}